\documentclass[11pt]{amsart}
\usepackage{amsmath}
\usepackage{hyperref}
\usepackage{mathtools}
\usepackage{amssymb}

\title{Fuzzy Logic and Markov Kernels}
\author{Rogier Brussee}
\address{Jheronimus Academy of Data Science, Sint Janssingel 92 5211 DA den Bosch/ Technical University Eindhoven}  
\email{R.Brussee@tue.nl, Rogier.Brussee@gmail.com}
\keywords{Fuzzy Logic, Markov Kernel, Probability Theory}
 
\date\today

\newcommand\R{\mathbb{R}}
\newcommand\N{\mathbb{N}}
\newcommand\B{\mathbb{B}}
\renewcommand\P{\mathbb{P}}

\newcommand\sigmaA{\mathcal{A}}
\newcommand\sigmaB{\mathcal{B}}

\newcommand{\powerset}{\mathcal{P}}

\mathchardef\mhyphen="2D
\newcommand\emdash{\mhyphen}

\newcommand\true{\mathbf{T}}
\newcommand\false{\mathbf{F}}
\newcommand\logicnot\neg
\newcommand\logicand\wedge
\newcommand\logicor\vee
\newcommand\logicalimplies{\mathbf{\to}}

\newcommand\fuzzynot\neg

\newcommand\fuzzyand[1]{\wedge_{#1}}
\newcommand\fuzzyandmin{\underline\wedge}
\newcommand\fuzzyandindep{\curlywedge}
\newcommand\fuzzyandmax{\overline\wedge}

\newcommand\fuzzyor[1]{\vee_{#1}}
\newcommand\fuzzyormin{\underline\vee}
\newcommand\fuzzyorindep{\curlyvee} 
\newcommand\fuzzyormax{\overline\vee}

\newcommand\fuzzyimpliesmin{\underline\logicalimplies}
\newcommand\fuzzyimpliesindep{\rightsquigarrow} 
\newcommand\fuzzyimpliesmax{\overline\logicalimplies}

\newcommand\ADD{\mathbf{add}}
\newcommand\AND{\mathbf{and}}
\newcommand\OR{\mathbf{or}}
\newcommand\NOT{\mathbf{not}}
\newcommand\IMPLIES{\mathbf{impl}}

\newcommand\qindep{q_{\mathrm{indep}}}
\newcommand\qmax{q_{\max}}
\newcommand\qmin{q_{\min}}

\newcommand\pfuzzy{p}

\newcommand\esssup{\operatorname{\mathrm{ess\, sup}}\limits}

\newcommand\union\cup

\theoremstyle{plain}

\newtheorem{definition}{Definition}

\newtheorem{example}{Example}

\newcommand\bibauthor[1]{{#1}}
\newcommand\bibtitle[1]{\textsl{#1}}
\newcommand\bibpublication[1]{{#1}}
\newcommand\bibpublisher[1]{{#1}}
\newcommand\bibyear[1]{{#1}}
\newcommand\bibpages[1]{{#1}}
\newcommand\bibvolume[1]{\textbf{#1}}
\newcommand\bibavailableat[1]{{\hfill \\ available at \texttt{#1}}}

\begin{document}

\maketitle
\begin{abstract}
Fuzzy logic is a way to argue with boolean predicates for which we only have a confidence value between 0 and 1 rather than a well defined truth value.
It is tempting to interpret such a confidence as a probability. We use Markov kernels, parametrised probability distributions, to do just that.
As a consequence we get general fuzzy logic connectives from probabilistic computations on products of the booleans, 
stressing the importance of joint confidence functions.
We discuss binary logic connectives in detail and recover the ``classic'' fuzzy connectives  as bounds for the confidence for general connectives. 
We push multivariable logic formulas as far as being able to define fuzzy quantifiers and estimate the confidence.
\end{abstract}

\section{Introduction}

The notion of fuzzy sets tries to formalise the idea that if we have a  set of things $X$ and a property $P$ for an element $x \in X$, (i.e. a predicate function $P:X \to \B$),  
we can express a level of confidence in the truth of $P(x)$, traditionally called ``belief'', as a function $p:X \to [0,1]$ . 
A predicate is in one to one correspondence with a subset $X_P=\{x\in X\mid P(x)\} \subset X$, so the belief function $p(x)$ can be interpreted as 
a level of confidence of membership of $x$ in $X_P$,  a ``fuzzy'' membership function.  
It is tempting to interpret this membership function as a probability. On second thought this is a bit problematic, however, since $p(x)$ is a function on $X$ and certainly not a 
probability distribution on $X$ as there is no way to  ``add up'' confidences at different points $x \in X$ to a confidence 1, and even for discrete spaces there is no reason why they should.    
Nonetheless the interpretation of fuzzy logic and its relation to probability theory has been a continued discussion since the inception of fuzzy logic
\cite{Gaines1978}\cite{Dubois-Prade1989}\cite{Dubois-Prade1993}\cite{lavioletteSeamanBarettDouglasWoodall1995}\cite{Zadeh1995}.
Here we will discuss a direct interpretation of fuzzy logic in terms of probability theory and more particularly in the Markov category. It interprets the belief function as a thinly disguised Markov kernel with values in the  Booleans, i.e.  instead of interpreting the belief function as a probability distribution \emph{on} $X$  we interpret the confidence function
 as defining a probability distribution on the \emph{booleans} which is  \emph{parametrised} by $X$.  Somewhat similar ideas seem to have been proposed by Loginov \cite{Loginov1966} based on conditional probability rather than Markov kernels.
 
The main contribution of this paper is that this interpretation gives a straightforward probabilistic interpretation of fuzzy logic and its properties, which become a direct consequence of the properties of traditional Boolean logic and standard probabilistic computations. 
In particular we immediately get a family of fuzzy connectives like ``and'' and ``or''  parametrised by the confidence that (n)either predicate is true. More conceptually, in this interpretation logic connectives are only properly defined if we have a \emph{joined} confidence in the  predicates  i.e. joined probability distribution on  two booleans 
rather than just the belief functions separately which only determines the marginals. 
This makes sense intuitively: even if we only have $50\%$ confidence in a predicate $P$, we must still have $0\%$ confidence in $P \logicand \logicnot P$ and can have $100\%$ confidence in $P \logicor \logicnot P$. 
Of course the strong correlation between the two predicates is reflected in the joint distribution of $(P, \logicnot P)$, but the joint confidence can also express more subtle ``if $P_1$ then $P_2$ is more or less likely'' correlations. 
The current approach makes it possible, to make use of joined confidences/beliefs for correlated predicates in a natural way.   
In fact, the same argument that gives  a fuzzy/probabilistic interpretation for binary logic relations also works for more general systems of logic formulas. 
We can even even extent this framework to quantification over infinite sets. 
However, while this Markov interpretation is useful and precise, it also requires to specify joint confidence functions which have $2^n -1$ degrees for $n$ predicates. 
Thus it is useful to have upper and lower bounds for logic connectives or results making assumptions like independence. 
It turns out that sensible bounds and independence simplifications exist and are given by connectives in their ``traditional'' fuzzy forms (plural!), 
``explaining'' why an approach with logic flavour often works. 
 
 The organisation of this paper is as follows: in section \ref{sec:MarkovKernels} we define Markov kernels and the Markov category, as a quick recap and to fix notation. In section \ref{sec:FuzzyLogic} we introduce Fuzzy logic, again mainly to fix notation and specify what form of fuzzy logic we use. In section \ref{sec:SystemsOfLogicExpressions} we reformulate the belief function in terms of Markov kernels and use them to define binary logic connectives. 
 As an application of this probabilistic interpretation, we describe the relation to "classical" logic connectives and prove a de Morgan Law. 
 We finally discuss the general case and quantifiers in section \ref{sec:SystemsOfLogicExpressions}. 
 
In the rest of this paper we use the standard notations  $\R, \N$ for the real numbers, and nonnegative integers, 
$[a,b]$, $(a,b]$ or $(a,b)$ for the closed or (half) open interval and the less standard $\B = \{\false, \true\}$ for the Booleans.

\subsection{competing interest}
The author gracefully acknowledges support from the RNOB Regio deal Noord Oost Brabant, but otherwise declares no competing interest in this research.

\section{Markov kernels}\label{sec:MarkovKernels}

A Markov kernel is a parametrised version of a probability distribution.   
Just like a probability distribution formalises a random value, a Markov kernel formalises a function with random values. 
The formal definition leans on the language of measure theoretic probability theory, but it is mainly used as solid foundation supporting an oiled machinery. 
The section is mainly needed because we liberally use the language from this section, in particular that of functoriality of pushforward.  
Some readers may prefer to only take a cursory look at the definition of Markov kernel
and pushforward, and go back to other definitions as needed. 
 
Let $X = (X, \sigmaA)$ and $Y = (Y, \sigmaB)$ be two measurable spaces, i.e. a set together with a $\sigma$-algebra of measurable subsets \cite[Ch 1]{Rudin1970}.

\begin{definition}
A  Markov kernel \cite[Appendix VI]{Doob1984} $F: X  \to Y$ is a map $\sigmaB \times X \to [0,1]$  such that
\begin{itemize}
\item for each $ B \in \sigmaB$,  the map 
\begin{align}
	F(B, \emdash):X &\to [0,1] \\
	                        x &\mapsto F( B, x ) 
\end{align}
is a measurable function on $X$,
\item for each $x \in X$ the map
\begin{align}
	F(\emdash, x): \sigmaB  &\to [0,1] \\
				         B &\mapsto F( B , x) 
\end{align}
is a probability measure on $Y$. 
\end{itemize}
\end{definition}
As a convenient (manifestly asymmetric!) notation we write $F(dy | x)$ for the probability measure on $Y$ defined by the point $x \in X$. 
We likewise write $F(B|x) = \int_B F(dy|x)$ for $F(B, x)$.  
The notation also suggests conditional probability. While this is not a bad way to think about it, the two concepts are related but  technically and conceptually slightly different: 
a conditional probability is a Markov kernel defined almost surely with respect to a probability measure on the ``universe of discourse'' $X$. This background measure on $X$ can become a block in interpretation, particularly if one liberally identifies measures with functions on discrete spaces. It is also the way people denote parametrised probability distribution, e.g. $G(d^n x | \Sigma, \mu)$ for a Gaussian distribution on $R^n$ that is parametrised by its covariance matrix $\Sigma$ and mean $\mu$.  

The $\sigma$-algebras and measurability of the function $F(B | \emdash)$ are a convenient technical setup allowing us to define integration without worries, and depending on the $\sigma$ algebra disallow pathological constructions using the axiom of choice (e.g. if $X$ and $Y$ are topological spaces and the $\sigma$-algebras are the corresponding Borel algebras, all continuous functions are allowed) or give built in adaptation to some ``pixelation'' of the space.   
The meat of the definition is the second part: it allows us to think of Markov kernels as probabilistic ``functions'' that assign a probability distribution $F(dy|x)$ on $Y$ rather than a specific value for every $x$ in $X$.
Conversely, every normal measurable function $f:X \to Y$ defines a Markov kernel\footnote{
If the $\sigma$-algebra $\sigmaB$ on $Y$ does not separate points, different functions may define the same kernel.}
 by
 $F(dy |x) = \delta_{f(x)}(dy)$ i.e. 
 \begin{equation}
 	F(B |x)  =   \begin{cases} 1 &\text{ if }  f(x) \in B\\ 0 &\text{ otherwise} \end{cases}
\end{equation} 
 In particular, the identity function on $X$ has the Markov kernel
 \begin{equation}
 	1_X(dx'|x) = \delta_x(dx') 
\end{equation}

Like functions, Markov kernels can be composed: if $F: X \to Y$ and $G:Y\to Z$ are Markov kernels, then the composition of the kernels $G\circ F$ is defined by 
\begin{equation}
	(G\circ F)(dz |x) = \int_Y G(dz |y)F(dy | x)
\end{equation}
i.e. for each measurable subset $C\subset Z$ and $x \in X$ we define
\begin{equation}
	(G\circ F)(C |x) = \int_Y G(C|y)F(dy|x)  
\end{equation}
One checks 
that $G\circ F$ is indeed a Markov kernel, that $\circ$ is associative and $1_X$ is the identity on $X$.
This gives measurable spaces and Markov kernels the structure of a category \cite[section I.2]{MacLane2013}, the Markov category \cite{Panangaden1999}, (aka Lawvere category \cite{Lawvere1962}, aka Giry monad\cite{Giry1982} aka the Stochastic Category \cite{Voevodsky2008}).
The category has an initial object, the empty set $\emptyset$, and a final object, the one point set $\star = \{\ast\}$. 
A probability measure $\P$ on a measurable set $X$ is ``the same thing'' as a Markov kernel $P:\star \to X$ because the Markov kernel is fully determined by the probability measure $P(dx|\ast) = \P(dx)$. 
Hence a probability space $(X, \sigmaA, \P)$  is ``the same thing'' as a pointed space $\star \to X$ in the Markov category and we will identify them.
More generally, if $X$ comes with a family of probability distributions parametrised by $S$, 
(i.e. a Markov kernel $\P: S \to X$)  then for every kernel $F:X \to Y$ we have a pushforward  $F_*\P = F \circ \P: S \to Y$. 
 The pushforward\footnote{%
 There is also a functorial contravariant pullback: a kernel $ \phi: Y \to T$ gives a kernel $F^*\phi: X \to T$ by $F^*\phi = \phi \circ F$ which satisfies $(G\circ F)^* = F^* \circ G^*$}: 
 is a covariant functor, i.e. for $G:Y \to Z$  we have 
 \begin{equation} 
 	(F\circ G)_* = F_*G_*
\end{equation}
This \emph{functoriality} is just expressing the fact that $\circ$ is associative here.   
If a kernel comes from a measurable function $f:X \to Y$ the pushforward is very explicit. For $B \in \sigmaB$
\begin{equation}
	(f_*P)(B | s)= P(f^{-1}(B) |s)
\end{equation}
i.e. it is just the pushforward of measures but parametrised with $s \in S$. 
Given $F_1: X \to Y_1 $ and $F_2: X \to Y_2 $ there does not, in general, exist a \emph{unique} Markov kernel 
\begin{equation}
	F_{12}: X \to Y_1 \times Y_2
\end{equation} 
with marginals $\pi_{1*}F_{12} = F_1$ and $\pi_{2*}F_{12} = F_2$, where $\pi_1, \pi_2$ are the projections on the first and second coordinate. 
This is  because marginal probability distributions do not, in general, determine a joint probability distribution even though the marginals may 
heavily restrict the joint distribution. 
We can, however, define a conditionally independent (i.e. independent for every $x$) joint kernel $F_1 \otimes F_2 : X \to Y_1 \otimes Y_2$ defined by the product measure \cite[Definition 7.7]{Rudin1970}
\begin{equation}\label{eq:IndependentMarkovKernels}
	(F_1 \otimes F_2)(d(y_1,y_2)|x) = F_1(dy_1|x)F_2(dy_2|x).
\end{equation}
On measurable subsets $B_1 \times B_2 \subset Y_1 \times Y_2$,  it is given by 
$(F_1 \otimes F_2)(B_1 \times B_2 | x) = F_1(B_1 |x) F_2(B_2 | x)$ which is then uniquely extended to the product $\sigma$-algebra using the Caratheodory extension theorem. 

To explain what we are going to do with logic connectives $\logicand$ and $\logicor$ it is useful to start with the more familiar situation of adding random variables. 
Therefore first assume that $(X, \sigmaA)$ is not just a measurable space but a probability space $(X, \sigmaA, \P)$ for some probability measure $\P$.
Recall that in this setup a random variable or stochast\footnote{%
In this paper uppercase ``end of alphabet'' latin letters denote (measure) spaces rather than stochasts.%
} 
is a measurable map $f:X \to \R$. 
The law of the stochast $f$  is the pushforward measure  $f_* \P$, which is a probability distribution on $\R$. 
Concretely, for an interval $(-\infty ,b] \subset \R$ the pushforward measure gives the cumulative distribution
 \begin{equation}
 	(f_*\P)((-\infty,b]) = \P(f^{-1}(-\infty,b]) = \P( f(x) \le b).
\end{equation}
For $F: X \to \R$ a Markov kernel, we have the pushforward probability measure $F_*\P = F\circ \P$ with a cumulative distribution   
\begin{equation}
	(F_*\P)((-\infty, b]) = \int_X F((-\infty,b] | x) \P(dx)
\end{equation}

For two stochasts $f_1, f_2: (X, \sigmaA, \P) \to \R$,  we have a well defined sum function $(f_1 + f_2)(x) = f_1(x) + f_2(x)$.  
The law of their sum  $f_1 + f_2$ is the probability measure $(f_1+f_2)_*\P$. However, it is well known that, in general, the law of the sum is \emph{not} determined by the laws ${f_1}_*\P$ and ${f_2}_* \P$ of the stochasts separately.  For example if  $f$  is normally distributed with mean $0$ and standard deviation $1$, and $f_1 = f$, $f_2 = f$, then $f_1$ and $f_2$ are also normally distributed with standard deviation $1$ and $f_1 + f_2 = 2f$ is normally distributed with standard deviation $2$, while if $f'_1 = f$, $f'_2 = -f$ then again $f_1$ and $f_2$ are again normally distributed with standard deviation $1$, but obviously $f_1 + f_2 = 0$. 
What \emph{does} determine the law of the sum is the \emph{joint} law $(f_1\times f_2)_*\P$ on $\R^2 = \R \times \R$. 
This follows from functoriality since if we define $\ADD(y_1, y_2) = y_1 + y_2$ then $f_1 + f_2 = \ADD\circ(f_1\times f_2)$ ,  and the joint law is 
\begin{equation}
	(f_1 + f_2)_*\P = (\ADD\circ (f_1 \times f_2))_*\P = \ADD_* \left((f_1\times f_2)_*\P\right).
\end{equation}

For Markov kernels $F_1,F_2 :X \to \R$ there is a very similar complication: they do not define a well defined kernel which one would reasonably call ``$F_1+F_2$''. 
In order to compute the distribution of the sum, we need a \emph{joint} Markov kernel 
i.e. a kernel $F_{12} : X \to \R^2$ with marginals ${\pi_1}_*F_{12} = F_1$ and ${\pi_2}_* F_{12} = F_2$, where again $\pi_1, \pi_2$ are the projections on the first and second coordinate. 
Such a joint kernel certainly exists since we can always assume take $F_1$ and $F_2$ are conditionally independent and take $F_1 \otimes F_2$

 are conditionally independent (i.e. independent for every $x$) and take $F_{12} = F_1 \otimes F_2$ on $\R^2$. 
 However, as mentioned earlier, even though $F_{12}$  is restricted by the marginals $F_1$ and $F_2$, there is, in general, no uniquely determined choice. 
\emph{Given} a joint Markov kernel, however, we have a Markov kernel that by abuse of notation we may call the sum
\begin{equation}
	F_1 +_{F_{12}} F_2 := \ADD_* F_{12}. 
\end{equation}
Here the notation $+_{F_{12}}$ is supposed to indicate that we use the \emph{joint} Markov kernel $F_{12}$ to define addition. 
This notation is a bit silly because $F_{12}$ already determines $F_1$ and $F_2$, but If no confusion seems possible we write
\begin{equation}\label{eq:suggestiveSumKernel}
	F_1 +_{12} F_2 := F_1 +_{F_{12}}  F_2.
\end{equation}
Since  $F_{12}$ although not uniquely determined by the marginals, may still be substantially restricted, and so $F_1 +_{12} F_2$ is restricted.

\begin{example}
Consider two Markov kernels $F_1, F_2: \Delta \to \R$ on the  interval $\Delta_1 = [0,1]$  with values in $\{\pm 1\} \subset \R$, representing the loss or gain for two parties in 
two Bernouilly experiments each with a chance $p \in \Delta$ for heads ($h$) and $(1-p )$ for tails ($t$). Party 1 wins on heads, Party 2 on tails. 
What we do not know is the correlation between the two Bernouilly experiments. We then have   
\begin{align}
	F_1 (dy_1 |p) &= (p \delta_{+1} + (1-p)\delta_{-1})(dy_1)\\
	F_2 (dy_2 |p) &= ((1-p) \delta_{+1}+ p \delta_{-1})(dy_2) 
\end{align}

An equivalent way to write down this kernel is as probabilities of the different outcomes
\begin{align}
	F_1( +1 |p) &= p,        &F_2( +1 |p) &= (1-p)  \\
	F_1( -1|p) &= 1- p       &F_2( -1|p) &= p      \\
	F_1( B |p) &= 0           &F_2( B |p) &= 0  \text{ if } \pm1 \notin B
\end{align}
where we used the usual abuse of notation in the probability distributions of discrete sets to  write  e.g. $F_1( +1 |p) = F_1(\{-1\}|p)$ which we will continue to do. 
The joint Markov kernels can then be parametrised as 
\begin{align}
	F_{12}( (+1, +1)|p) &= p_{ht},    &F_{12}( (+1, -1) |p) &= p_{hh}  \\
	F_{12}( (-1, +1) |p) &= p_{tt}     &F_{12}( (-1, -1) |p) &= p_{th}    \\
	F_{12}( B          |p) &= 0         \text{ if } (\pm1, \pm1) \notin B
\end{align}
where the condition that $F_{12}$ has $F_1$ and $F_2$ as marginals gives
\begin{align}
	p_{hh} + p_{ht}  &= p         &p _{hh} + p_{th}  &= p \\
	p_{th} + p_{tt}   &= 1- p      &p_{ ht} + p_{tt}  &= 1- p
\end{align}
with the inequalities $p_{ab} \ge 0$. The system of equations is equivalent to 
\begin{align}
	p_{hh} + p_{ht}  +  p_{th} + p_{tt}  &= 1  \\
	p_{hh} + p_{ht}                              &= p  \\
	              p_{ht}  +                 p_{tt} &= 1-p    
\end{align}
and the solutions are of the form
\begin{align}
	p_{hh} &=   q                    & p_{ht} &=   p - q \\
	p_{th} &=    p - q               & p_ {tt} &=   1 - 2p + q   
\end{align}
with $\min(2p -1,  0) \le q \le p$.   
Given this joint kernel, the sum kernel is 
\begin{align}
	(F_1 +_{12} F_2)( +2  | p) &= p_{ht} = p - q \\
	(F_1 +_{12} F_2)(  0   | p) &= p_{hh} + p_{tt} = 1 - 2p + 2q \\
	(F_1 +_{12} F_2)( -2  | p) &= q_{th} = p - q 
\end{align}
\end{example}
 What we are going to do for fuzzy logic  is much the same except rather than adding numbers, we are going to apply  ``and'' or ``or'' to boolean values. 
 
\section{Fuzzy Logic}\label{sec:FuzzyLogic}

For a set $X$, given the belief (confidence) $p:X \to [0,1]$  in a predicate $P: X \to \B$  the belief in $\logicnot P(x)$ is $1 - p(x)$.
\begin{equation}
	(\logicnot p) = 1 - p
\end{equation}

For two properties $P_1, P_2$ with belief functions $p_1, p_2$, we consider the belief (confidence) in  $P_1(x) \logicand P_2(x)$ or equivalently  the belief (confidence) in  
 $x \in X_{P_1} \cap X_{P_2}$ as a function of $x$.   Several natural choices for a fuzzy ``and'' that only depend on the values $p_1(x)$ and $p_2(x)$ have been proposed 
\cite{AlsinaTrillasValverde1983}, in particular: 
\begin{align}
        	p_1 \fuzzyandmin    p_2  &= \max(0, p_1 + p_2  -1)  \label{eq:andmin} \\
	p_1 \fuzzyandindep  p_2 &=  p_1p_2                         \label{eq:andindep} \\
	p_1 \fuzzyandmax   p_2 &=  \min(p_1, p_2).             \label{eq:andmax}
 \end{align}	
 Note that all choices reduce to the usual ``and'' connective $p_1 \logicand p_2$, if $p_1$ and $p_2$ only take the value $0$ or $1$ and that 
 \begin{equation}
 	p \fuzzyandmin q \le  p \fuzzyandindep q \le p \fuzzyandmax  q
\end{equation} 
Likewise consider the belief (confidence) in  $P_1 (x) \logicor P_2(x)$ i.e. the confidence in  $x \in X_{P_1} \cup X_{P_2}$ for which 
\begin{align}
	p_1 \fuzzyormin p_2 &= \max(p_1, p_2)                 \label{eq:ormin}\\
	p_1 \fuzzyorindep p_2 &= p_1 + p_2 - p_1p_2       \label{eq:orindep}\\
	p_1 \fuzzyormax p_2 &= \min(1, p_1 + p_2) .        \label{eq:ormax}
\end{align}
has been proposed. 
Note again that all choices reduce to the usual ``or'' connective $p_1 \logicor p_2$ if $p_1$ and $p_2$ only take the values $0$ or $1$ and that  
 \begin{equation}
 	p_1 \fuzzyormin p_2 \le p_1 \fuzzyorindep p_2 \le p_1 \fuzzyormax p_2.
\end{equation}
We will see later that $\fuzzyandmin$ and $\fuzzyandmax$ are extremals of a 1 parameter family of such connectives with $\fuzzyandindep$ as an intermediate value, 
and similarly for $\fuzzyormin$, $\fuzzyormax$ and $\fuzzyorindep$.

Given the proposed fuzzy ``and'' and ``or connectives, the classical de Morgan's law
\begin{equation}\label{eq:deMorganClassic}
 	 \logicnot(a \logicor b) = (\logicnot a) \logicand (\logicnot b), \ \forall a,b \in \B
\end{equation} 
for Boolean truth values have fuzzy analogues 
\begin{align}
	p_1 \fuzzyormin p_2   
	                                    &= \logicnot ((\logicnot p_1)\fuzzyandmax (\logicnot p_2))     \label{eq:deMorganMin}\\
	p_1 \fuzzyorindep p_2 
					   &= \logicnot((\logicnot p_1) \fuzzyandindep (\logicnot p_2))  \label{eq:deMorganIndep}\\
	p_1 \fuzzyormax p_2   
					   &= \logicnot((\logicnot p_1) \fuzzyandmin(\logicnot p_2))     \label{eq:deMorganMax}
\end{align}
we will later see that there is a De Morgan law for a whole 1 parameter family with formulas \eqref{eq:deMorganMin}, \eqref{eq:deMorganIndep} and \eqref{eq:deMorganMax}
as special cases.                                 

\section{Fuzzy logic connectives and  Markov kernels}\label{sec:FuzzyLogicAndMarkovKernels}

With the notion of Markov kernel at hand, we can interpret a fuzzy set or fuzzy predicates on a measurable set  $(X, \mathcal A)$  as a predicate kernel  $ P:  X \to \B$.  
A belief function $p:X \to [0,1]$ as in section \ref{sec:FuzzyLogic} is equivalent to a predicate kernel with 
\begin{align}\label{eq:BeliefP}
	P(\true | x)  &=  p(x) \\
	P(\false |x)  &= 1 - p(x). 
\end{align}
Conversely the kernel  predicate defines a belief function in this way.   Again we  abused notation in the probability distributions of discrete sets to write  
e.g. $P(\true|x)$ for $ P(\{\true\}|x)$ here. 
This interpretation will allow us to think of fuzzy logic as a ``shadow'' of ordinary logic by suitably applying functoriality. In the process we will quickly find the need to use joint confidence functions or joint kernels. As explained in section \ref{sec:MarkovKernels}, the introduction of a $\sigma$-algebra of subsets and assuming that $p$ is measurable is a technical detail. In fact for most of what follows we could take $\sigmaA = \powerset(X)$ the full powersetof X. 

\subsection{logical connective ``not''}

As a warmup, consider the belief function of the ``not'' connective $\logicnot$.  There is an associated function
\begin{align}
	\NOT: \B \to \B\\
		  t \mapsto \logicnot t
\end{align}
The Markov kernel for $\logicnot P : X \to \B$ is then defined functorially on predicate kernels
\begin{equation}
	(\logicnot P) := (\NOT_*P).
\end{equation}
Its corresponding belief function is 
\begin{align}
	(\logicnot p)(x) &=(\logicnot P)(\true |x) \\
	                          &=  (\NOT_*P)(\true |x)   \\
	                          &= P(\NOT^{-1}(\true) |x)  \\
	                          &= P(\{\false\} |x)  \\
	                          &= 1 - P(\true |x)  \\
	                          &= 1- p(x)
\end{align}  
just as expected. 

\subsection{logical connective ``and''}

Now, just as the sum of two  Markov kernels with values in $\R$ turned out not to be well defined without a joint kernel, to define a logical ``and'' connective 
of predicate Markov kernels $P_1$ and $P_2$ (Markov predicates for short), we need a \emph{joint} Markov predicate  $P_{12}$ lifting $P_1$ and $P_2$. Spelled out: 
given Markov predicates  $P_1, P_2: X \to \B$ uniquely defined by  $P_i(\true|x)  = p_i(x)$, we need a joint Markov predicate $P_{12}: X \to \B^2$  
with conditional marginals  $\pi_{1*} P_{12} = P_1$ and  $\pi_{2*}P_{12} = P_2$ where $\pi_i$ is projection onto the $i$-th coordinate.
Again, such a lift exists because again as a special case of \eqref{eq:IndependentMarkovKernels} we can take the joint Markov predicate  $P_{12} = P_1\otimes P_2$ 
assuming independence of $P_1$ and $P_2$. Here this becomes very explicit:   
\begin{equation}\label{eq:IndependentKernels}
	(P_1\otimes P_2)((a,b)|x) = P_1(a |x)P_2(b|x).
\end{equation}
However, since $\B^2$ is small we can easily write down the general Markov predicate lifting $P_1$ and $P_2$:  
\begin{align}
	P_{12}((\false, \false) |x) &= p_{FF}(x)   &P_{12}((\false, \true) |x) &= p_{FT} (x) \\
	P_{12}((\true , \false)  |x) &= p_{TF}(x)  & P_{12}((\true , \true) |x) & = p_{TT}(x) 
\end{align}
where the $p_{ab}(x) \ge 0$ for all $x$ and 
\begin{align}
	p_{FF}(x) + p_{FT}(x)  &= 1- p_1(x)    &p _{FF}(x) + p_{TF}(x)  &= 1-p_2(x) \\
	p_{TF}(x) + p_{TT}(x)  &= p_1(x)        &p_{FT}(x) + p_{TT}(x)   &= p_2(x).
\end{align}
These equations are equivalent to 
\begin{align}
	p_{FF}(x) + p_{FT}(x)  +  p_{TF}(x) + p_{TT}(x)  &= 1  \\
	                                        p_{TF}(x) +  p_{TT}(x)  &=  p_1(x)  \\
	                   p_{FT}(x)  +                     p_{TT}(x)   &=  p_2(x)    
\end{align}
which has solutions
\begin{align}
	p_{FF}(x) &= q(x)                           & p_{TF} &= (1 - p_2(x)) - q(x)                   \label{eq:lift2a} \\
	p_{FT}(x) &= (1-p_1(x))-q(x)          & p_{TT} &= p_1(x) + p_2(x) - 1  + q(x).    \label{eq:lift2b}
\end{align}
Thus in this parametrisation (there can be others) the joint Markov predicate is determined by  a ``belief function'' $q(x)$  for \emph{both} predicates to be wrong. 
The inequalities $p_{ab}(x) \ge 0$ translate inequalities  
\begin{equation}\label{eq:q-inequalities}
	0 \le \qmin(x) \le  q(x) \le \qmax(x) \le 1
\end{equation}
for $q(x)$ where 
\begin{align}
	 \qmin(x) &:=  \max(0, 1 - (p_1(x) + p_2(x)))          \label{eq:qmin}\\
	 \qmax(x) &:= 
	                       1 - \max(p_1(x), p_2(x)).                 \label{eq:qmax}
\end{align}
We also define
\begin{equation} \label{eq:qindep}
	\qindep(x) :=  (1 -p_1(x))(1-p_2(x)).     
 \end{equation}

Given the joint Markov predicate we can now \emph{by abuse of notation} define a Markov predicate $P_1 \fuzzyand{12} P_2$ for the ``and'' connective by functoriality
\begin{align}\label{eq:suggestiveAndMarkovPredicate}
	P_1 \fuzzyand{12} P_2 := P_1 \fuzzyand{P_{12}} P_2 :=  \AND_*P_{12}.
\end{align}
As for the sum kernel in \ref{eq:suggestiveSumKernel},  the notation is a bit silly because $P_1$ and $P_2$ are already determined by $P_{12}$ while $P_1$ and $P_2$ do not determine the joint Markov predicate $P_{12}$.
It is completely determined by its belief function
\begin{align}
	(P_1 \fuzzyand{12} P_2)(\true |x) &:= (\AND_*P_{12})(\true | x) \\
	                                    &= P_{12}(\AND^{-1}(\true) |x) \\
	                                    &= P_{12}(\{(\true, \true)\} | x) \\
	                                    &= p_{TT}(x) \\
	                                    &=p_1(x) + p_2(x)  +  q(x) -  1                            
\end{align}  
i.e. the belief function for the fuzzy ``and'' defined by the joint kernel is 
\begin{equation}
	p_1(x) \fuzzyand{q(x)} p_2(x) :=  (P_1 \fuzzyand{12} P_2)(\true |x) = p_1(x) + p_2(x)  +  q(x) -  1. 
\end{equation}
By the inequalities \eqref{eq:q-inequalities} for $q(x$ we get
\begin{equation}
	\max(0, p_1(x) + p_2(x) -1)   \le p_{TT}(x)  \le \min(p_1(x), p_2(x)) 
\end{equation}
and these inequalities are sharp if $q(x) = \qmin(x)$ (for the lower bound), or $q(x) = \qmax(x)$  (for the upperbound).
By definitions \eqref{eq:andmin}, \eqref{eq:andmax}
\begin{align}
	p_1(x) \fuzzyandmin p_2(x) &= p_1(x) \fuzzyand{\qmin} p_2(x) \\
	p_1(x) \fuzzyandmax p_2(x) &= p_2(x) \fuzzyand{\qmax} p_2(x)
\end{align}
and so
\begin{equation}
	p_1(x) \fuzzyandmin p_2(x) \le p_1(x) \fuzzyand{q(x)} p_2(x) \le p_1 \fuzzyandmax p_2(x) .	
\end{equation}

In case $P_1$ and $P_2$ are conditionally independent and $P_{12} = P_1 \otimes P_2$ we have
\begin{align}
	p_{FF}(x) &= (1 -p_1(x))(1-p_2(x))  = \qindep \\
	p_{TT}(x) &= p_1(x)p_2(x) = p_1(x) + p_2(x) + \qindep -1
\end{align}
hence comparing with \eqref{eq:andindep} we find
\begin{equation}
 	p_1(x) \fuzzyandindep p_2(x) = p_1(x) \fuzzyand{\qindep} p_2(x)
\end{equation}
 
 \subsection{logical connective ``or''}
 
 We can proceed similarly for the ``or'' connective. Using the same abuse of notation as for ``and''  we define a Markov predicate
 \begin{equation}
    P_1 \fuzzyor{12} P_2  := P_1 \fuzzyor{P_{12}} P_2 := \OR_*P_{12}
  \end{equation}
 with  belief function $p_1(x) \fuzzyor{q(x)} p_2(x) := P_1 \fuzzyor{P_{12}} P_2(\true |x)$. We then find
\begin{equation}
	 p_1(x) \fuzzyor{q(x)} p_2(x) := (\OR_*P_{12})(\true | x) 
	                                       = 1 - q(x).
\end{equation} 
By the inequalities \eqref{eq:q-inequalities} for $q(x)$ we get inequalities
\begin{equation}\label{eq:orbounds}
	 \max(p_1, p_2) \le  1 - q(x) \le 
	                                                   \min(1, p_1(x) + p_2(x))
\end{equation}
which are sharp if $q(x) = \qmax(x)$ (for the lowerbound) and  $q = \qmin$ (for the upperbound).
By  \eqref{eq:ormin} and \eqref{eq:ormax} the inequalities \eqref{eq:orbounds} then translate to
\begin{equation}	
	p_1(x) \fuzzyormin p_2(x) \le p_1(x) \fuzzyor{q(x)} p_2(x) \le  p_1(x) \fuzzyormax p_2(x).
\end{equation}
Finally for the conditionally independent case $P_{12} = P_1 \otimes P_2$ we have
\begin{equation}
	           (\OR_*(P_1 \otimes P_2))(\true|x)  =  1 - (1-p_1)(1-p_2) = p_1(x) + p_2(x) - p_1(x)p_2(x),
\end{equation}
and by  \eqref{eq:orindep} and \eqref{eq:qindep} we find
\begin{equation}
	p_1(x) \fuzzyor{\qindep} p_2(x) = p_1(x) \fuzzyorindep p_2(x) 
\end{equation}	 	 

\subsection{logical implication}

We can do exactly the same with other logic connectives in particular implication. For $a,b \in \B$ we have 
\begin{equation}
	\IMPLIES(a,b) = a \logicalimplies b = \logicnot a \logicor b = \logicnot( a \logicand \logicnot b).
\end{equation}
We therefore define a Markov predicate
\begin{equation}
	P_1 \logicalimplies_{12} P_2 := P_1 \logicalimplies_{P_{12}} P_2  := \IMPLIES_*P_{12}
\end{equation}
with a belief function
\begin{equation}
	p_1(x) \logicalimplies_{q(x)} p_2(x)  = \IMPLIES_*P_{12}(\true |x)  \\
	             = 1 - P_{TF}(x) 
	             = p_2(x) + q(x).
\end{equation}
Now defining
\begin{align}
	p_1(x) \fuzzyimpliesmin p_2(x)  &=   p_2(x) + \qmin(x)   
	                                                     = \max(p_2(x), 1 - p_1(x))\\
	p_1(x)\fuzzyimpliesindep p_2(x)  &=   p_2(x) + \qindep(x)
	                                                       = 1 - p_1(x) + p_1(x)p_2(x) \\
	p_1(x) \fuzzyimpliesmax p_2(x) &=   p_2(x) + \qmax(x)  
	                                                     = \min(1, 1 + p_2(x) - p_1(x)) 
\end{align}
we have 
\begin{equation}
	p_1(x) \fuzzyimpliesmin p_2(x) \le p_1(x) \logicalimplies_{q(x)} p_2(x) \le p_1(x) \fuzzyimpliesmax p_2(x)
\end{equation}

\subsection{de Morgan's law}

It is easy to prove a general fuzzy de Morgan's law using direct computation. We will also give a more conceptual proof  showing it is a 
consequence of the classical de Morgan's law \eqref{eq:deMorganClassic} and functoriality. 
We have the identity
\begin{equation}
	q = (1 - p_1)  + (1 - p_2)  +   (p_1 + p_2 + q  - 1 ) - 1 = (\logicnot p_1) + (\logicnot p_2) + (p_1 \fuzzyand{q} p_2 ) -1. 
\end{equation}
In other words, the general fuzzy analogue of the de Morgan's law \eqref{eq:deMorganClassic} is
\begin{equation}\label{eq:nor=andnn}
	\logicnot\left(p_1(x) \fuzzyor{q(x)} p_2(x)\right)  =   (\logicnot p_1(x)) \fuzzyand{ (p_1(x) \fuzzyand{q(x)} p_2(x) )} (\logicnot p_2(x))
\end{equation}
or 
\begin{equation}\label{eq:deMorganFuzzyBelieve}
	    p_1(x) \fuzzyor{q(x)} p_2(x)   =   \logicnot\left((\logicnot p_1(x)) \fuzzyand{ (p_1(x) \fuzzyand{q(x)} p_2(x) )} (\logicnot p_2 (x))\right).
\end{equation}
	                                        
For a more conceptual derivation of  \eqref{eq:nor=andnn} rewrite the classic Boolean the Morgan's law \eqref{eq:deMorganClassic} as
\begin{equation}
	\NOT\circ \OR = \AND \circ (\NOT \times \NOT).
\end{equation}
Then by functoriality
\begin{align}
	\NOT_*\OR_* &= (\NOT\circ\OR)_*\\
	                       &= (\AND \circ (\NOT \times \NOT))_* \\
	                        &= \AND_* (\NOT \times \NOT)_* .
\end{align}
Thus we get 
\begin{equation}\label{eq:norkernel=andnnkernel}
	\NOT_*(\OR_*P_{12}) = \AND_* \left((\NOT \times \NOT)_* P_{12}\right),
\end{equation}
and  so 
\begin{equation} \label{eq:and-NNkernel}
	(\NOT_* (\OR_*P_{12}))(\true |x) 
	                                     = \AND_*\left((\NOT \times \NOT)_* P_{12}\right)(\true | x) 
\end{equation}
But the ``q'' of the joint kernel $(\NOT \times \NOT)_* P_{12}$  is 
\begin{equation}
	((\NOT \times \NOT)_* P_{12}) (\false, \false) |x) = P_{12}((\true, \true) |x) 
	                                                                             =  p_1(x) \fuzzyand{q(x)} p_2(x)  
\end{equation}
whereas the marginals ``$p_1$'' and ``$p_2$'' of $(\NOT \times \NOT)_* P_{12}$ are 
\begin{align}
	\left(\pi_{i*}(\NOT\times\NOT)_*P_{12})\right(\true|x)   &= \left(\NOT_*\pi_{i*}P_{12}\right)(\true |x) \\
	       &= (\NOT_* P_i)(\true|x) \\
	       &= \logicnot p_i(x)
\end{align} 
Thus writing the left and right hand side of \eqref{eq:and-NNkernel} in terms of belief functions we get the Morgan's law \eqref{eq:nor=andnn}.

\section{Systems of logic expressions}\label{sec:SystemsOfLogicExpressions}

It should now be clear that in order to use the Markov predicate interpretation to define fuzzy logic predicates in multiple variables, we require, in general, a  joint distribution probability distributions for the logic variables. 
A set of general logic formulas can be considered as a map $L:\B^n \to \B^m$, $0\le n, m < \infty$. In fact a single map $L_0 : \B^n \to \B$ can always be considered as evaluating an element $\tilde L_0$ in $\B[t_1,...,t_n]$, the free boolean algebra on $n$ letters. For notational simplicity consider the $n = 2$ case, the general case being similar. 
We can define
\begin{equation}
	\tilde L_0(t_1, t_2) = L_0(\true,\true)t_1t_2  \logicor 
	                                L_0(\false,\true)\bar t_1 t_2 \logicor
	                                L_0(\true,\false)t_1\bar t_2 \logicor
	                                L_0(\false,\false) \bar t_1\bar t_2 
\end{equation} 
where juxtaposition is $\logicand$ and $\bar t = \logicnot t$.  Since every logic formula in $\B[t_1,...,t_n]$ 
can be written uniquely in normal form as monomials of $t$'s and $\bar t$'s ``or''ed together, 
we see that the map $L_0$ defines $\tilde L_0$ and vice versa. We will therefore no longer make a notational difference.  

Suppose then that we are given \emph{joint} Markov predicates  $P_{\{1,\ldots, n\}}:X \to \B^n$ or equivalently
a full set of $2^n$ \emph{joint} belief (confidence) functions 
\begin{equation}
	p_{a_1\ldots a_n}(x) = P_{\{1,\ldots, n\}}((a_1\ldots a_n) | x)
\end{equation}
one for each ``bitstring'' $(a_1, \ldots a_m) \in \B^n$. They are subject to the usual conditions for a probability distribution on a discrete set
\begin{equation} 
	\sum_{(a_1, \ldots a_n)}  p_{a_1, \ldots a_n}(x) = 1  \text{ and } p_{a_1, \ldots a_n}(x) \ge 0 \  \forall (a_1,\ldots, a_n) \in \B^n. 
\end{equation} 
Thus, we have $2^n -1$ degrees of freedom subject to $n$ conditions, 
one for each of the believe (confidence) functions $p_1(x)  \ldots p_n(x)$ from the the marginals $P_i = \pi_{i*} P_{\{1,...n\}}$. 
Of course one can also consider intermediate marginals, e.g. for every pair. By the the above $n = 2$ case, these are defined by the belief functions $p_i(x)$ and for every pair, the belief
\begin{equation}
	q_{ij}  = \pi_{ij*}P_{\{1,...n\}}((\false, \false)|x) = q_{ji}
\end{equation}
that neither is true, giving  $n + n(n-1)/2$ conditions. 
\footnote{Further assumptions that may be natural depending on the application, may reduce the degrees of freedom further. 
E.g. if we can assume conditional independence for $P_1, \ldots, P_n$ we have $p_{a_1\ldots a_n}(x) = \prod_{i = 1}^n p_{a_i}(x)$. 
Invariance of the joined Markov predicate under permutations ensures that $p_{a_1\ldots a_n}$ only depends on the number of $\true$'s in the ``bitstring'' $a_1, \ldots, a_n$, 
giving $(n+1) -1 = n$ degrees of freedom, with 1 condition for the marginal $p_1(x) =  \cdots = p_n(x)$.}
    
In any case, by functoriality, given a a system of joint Markov predicates  $P_{\{1,\ldots, n\}}:X \to \B^n$ and a system of logic formulas $L: \B^n \to B^m$ 
we can define fuzzy version of $L$ applied to $P_{\{1,\ldots, n\}}$ as a joint Markov predicate $X \to \B^m$
\begin{equation}
	 L_{1\ldots n} (P_1, \ldots P_n) := L_{P_{1 \ldots n}}(P_1, \ldots P_n)  := L_* P_{1 \ldots n}
\end{equation} 
Again it is a silly \emph{abuse of notation} to pretend it is a fuzzy logic predicate of the Markov predicates $P_1, \ldots P_n$ separately.
The Markov predicate is completely determined by the joint belief (confidence) functions
\begin{equation}
	p_{L, a_1\ldots a_m}(x) = (L_*P_{\{1,\ldots, m\}})((a_1\ldots a_m) | x)
\end{equation}
one for each ``bitstring'' $(a_1, \ldots a_m) \in \B^m$, which are likewise subject to the conditions for a probability distribution  $\sum_{(a_1, \ldots a_m)}  p_{L, a_1, \ldots a_m}(x) = 1$ and $p_{L, a_1, \ldots a_m}(x) \ge 0$.

\subsection{Fuzzy quantification}

The general setup assuming $n, m$ finite, does still not capture all natural examples. In particular we want to use the logical quantifiers  $\exists_{i \in I}P_I = \logicor_{i \in I} P_i$and $\forall_{i \in I} P_i = \logicand_{i \in I} P_i$.
 In the case of finite sets this is covered by the above, but already in the case of countable sets we have an extension. However in the countable case quantifiers  still define a map $B^\N \to \B$.  
The space $\B^\N$ is a (non trivial) measurable space \footnote{the measurable sets are the $\sigma$-algebra generated by the inverse images of subsets of $\B^M$, $ M < \infty$ under the projection $\B^\N \to B^M$} and we study measurable ``logical connectives'': $\B^\N \to \B$. 
Unsurprisingly, $\exists$ and $\forall = \logicnot \circ \exists \circ (\logicnot^\N)$ are measurable. 
We will only discuss the $\exists$ quantifier to which the $\forall$ quantifier can be reduced using the Morgan's law or treated similarly, \emph{mutatis mutandis}. See also 
\cite{LiuKerre1998}
Our goal will be to derive sensible lower and upper bounds for its belief function that they only require finite marginals, and are therefore easier to handle.

Consider the countable infinite case first
\begin{equation}
	\OR_\N(t_1,t_2, \ldots) =\logicor_{i \in \N}. t_i
\end{equation}
with for each $I \subset \N$ a counterparts $\OR_I((t_i)_{i\in I}  = \logicor_{i\in I}  t_i$. 
A joint Markov predicate $P_\N: X \to \B^\N$ gives marginals  $P_I: X \to \B^I$ and for $I_1 \subseteq I_2  \subseteq \N$
\begin{align}
	(\OR_{I_1*} P_{I_1})(\true|x) &=  1 - P_{I_1}((\false | x)_{i \in I_1}) \\
	                                                &\le 1 - P_{I_2}((\false|x)_{i \in I_2} )  \\
	                                                &= (\OR_{I_2*}P_{I_2})(\true|x) .
\end{align}	   
In particular from the subsets  $I = \{i\}$ we get the lower bound.
\begin{equation}\label{eq:supLowerBound}
	\sup_i \P_i(\true | x) =  \sup_{i} p_i \le (\OR_{\N*} \P_\N)(\true|x) 
\end{equation}
and for $I= \{i,j\}$ the lowerbound
\begin{equation}\label{eq:QijLowerBound}
	\sup_{ij}  (\OR_{ij*} P_{ij})(\true|x)  
	                 = \sup_{ij} p_i(x) \fuzzyor{q_{ij}(x)} p_j(x) 
	                 \le (\OR_{\N*}\P_\N)(\true|x)  
\end{equation}

A sensible upper bound is a little harder to get by. Of course we have the trivial bound
\begin{equation}
	(\OR_{\N*}\P_\N)(\true) \le 1
\end{equation}
We get potentially better bounds as follows: Define 
\begin{equation}
	C_I = \{ t_i = \true \text{ for some } i \in I \} \subset \B^\N.
\end{equation} 
Starting from a disjoint partition $\N = \coprod_n  I_n$ we then have
\begin{equation} 
	\B^\N = \{(\false, \false, \ldots)\} \coprod \union_n C_{I_n} 
\end{equation}
and therefore
\begin{align}
	1 = P_\N(\B^\N | x) &\le \P_\N((\false, \false, \ldots) |x ) + \sum_n P_\N(C_{I_n} |x) \\
	                               & =  \P_\N((\false, \false, \ldots)|x ) + \sum_n P_{I_n}(C_{I_n} |x)
\end{align}
where we used $C_{I_n}$ for both $\{ t_i = \true \text{ for some } i \in I_n \} \subset \B^{I_n}\}$ and $\{ t_i = \true \text{ for some } i \in I_n \} \subset \B^\N\}$. 
We then have 
\begin{equation}
	(\OR_{\N*}P_\N)(\true |x) \le \sum_n (\OR_{I_n*}P_{I_n})(\true |x)
\end{equation}
in particular for $I_n = \{n\}$
\begin{equation}\label{eq:sumUpperbound}
	(\OR_{\N*}P_\N)(\true |x) \le \sum_n \P_{\{n\}}(\true|x) = \sum_n p_n(x)
\end{equation}
and for $I_n = \{i_n, j_n\}$
\begin{equation}\label{eq:sumQijUpperbound}
	(\OR_{\N*} P)(\true|x) \le \inf_{\N = \union_n \{i_n, j_n\}} \sum_n  p_{i_n}(x) \fuzzyor{q_{i_nj_n}(x)}  p_{j_n}(x)
\end{equation}
Of course these bounds can be rather poor. Better bounds can be obtained from decomposing $\B^\N$ in disjoint pieces.
E.g. we can decompose as
\begin{equation}
	\B^\N = \{(\false, \false,\ldots)\} \coprod C_1 \coprod  \coprod_{n \ge 2} C_n \setminus C_{n-1}.
\end{equation} 
However we can then no longer reduce to marginal joint Markov predicates with a finite number of predicates
although additional assumptions that give  Markov predicate for predicate $n$ conditional on the value of predicates $1, \ldots n-1$ determine the $P_\N(C_n \setminus C_{n+1}|x)$, 
but we will not pursue this further.
Finally the finite marginals do determine the infinite ``or'': defining $C_M = C_{\{1, \ldots, M\}}$, we have   
\begin{equation}
	C_\N  = \union_{M}  C_M
\end{equation}
and therefore
\begin{equation}
	(\OR_{\N*}P_\N)(\true |x) = P_\N(C_\N|x)  = \lim_{M \to \infty} P_\N(C_M |x) = \lim_{M\to \infty} (\OR_{M*}P_{\{1\ldots, M\}})(\true |x)
\end{equation}
by the monotone convergence theorem.
 
 Where this countably infinite case shows up is in trying to quantify over the (not necessarily countable) set $X$, i.e. making fuzzy sense of a classical predicate expression like 
\begin{equation}
	\exists x \in X : P(x) = \logicor_{x \in X} P(x)
\end{equation}
From the above, for a sensible fuzzy definition we should start with a lift to joint Markov predicate that lifts the (possibly uncountably many) $P(\true |x)_{x \in X}$. 
 Since $P(\true |x)$ is a measurable function of $x$, it is a the supremum of a countable sequence of simple functions taking on a finite number of values. Thus we can  taking on a finite number of values, and countable limit of a function taking on on finite number of values. This suggests that we should consider a lift 
\begin{equation}
	P_\N: X^\N \to \B^\N
\end{equation}
such that
\begin{equation}
	(\pi_{i*}P_\N)(\true | (x_1, x_2, \ldots, ))  =  P_\N(\{t_i = \true\} | (x_1, \ldots x_n) ) = P( \true | x_i)
\end{equation}
Again such a lift can be constructed using a conditional independence construction, but is not necessarily unique.    
We can then follow the above and define the most general fuzzy version of the $\exists$ quantification as 
\begin{equation}
	\pfuzzy_{\exists x:P(x)} = \sup_{(x_1, x_2, \ldots ) \in X^\N} (\OR_{\N*}P_\N)(\true | (x_1, x_2, \ldots))
\end{equation}
We can now play the game of getting useful bounds by going to marginals. 
The case $n=1$ gives 
\begin{align} 
	\sup_{x, \in X} p(x) = \sup_{(x, x_2, \ldots)} (\pi_{1*}P_\N)(\true | (x, x_2, \ldots )) &\le  \pfuzzy_{\exists x:P(x)}                  
\label{eq:supLowerboundLocal}\\
	\sum_{x \in X}  p(x) = \sup_{(x_1, x_2, \ldots)} \sum_{i} (\pi_{i*}P_\N)(\true | (x_1, x_2, \ldots)) & \ge  \pfuzzy_{\exists x:P(x)}     
\label{eq:sumUpperboundLocal}
\end{align}
The  $n = 2$ case is more interesting. 
Spelled out,  It means that in addition to the belief function $p(x)$ we have a measurable function $q(x_1, x_2):  X^2 \to [0,1]$ such that $ \qmin(p(x_1), p(x_2)) \le q(x_1, x_2) \le \qmax(p(x_1), p(x_2))$  which together with the belief function $p(x)$ defines a lift $P_{1,2}: X^2 \to \B^2$ such that (compare \eqref{eq:lift2a} \eqref{eq:lift2b})
\begin{align}
	p_{FF}(x_1, x_2) &= q(x_1, x_2)            & p_{FT}(x_1,x_2) &=  1 - p(x_2) - q(x_1, x_2) \\
	P_{TF}(x_1,x_2)    &= 1 - p(x_1) - q(x_1, x_2) & p_{TT}(x_1, x_2) &= p(x_1) + p(x_2) - 1 - q(x_1, x_2). 
\end{align}
We then have upper and lower bounds
\begin{align}
	 \sup_{(x_1, x_2) \in X^2} p(x_1) \fuzzyor{q(x_1, x_2)} p(x_2) &\le \pfuzzy_{\exists x:P(x)} \\
	\sup_{(x_1, x_2, \ldots) \in X^\N} \inf_{\N = \union_n \{i_n, j_n\}}  \sum_n p(x_{i_n}) \fuzzyor{q(x_{i_n}, x_{j_n})} p(x_{j_n}) &\ge \pfuzzy_{\exists x:P(x)}
\end{align}

A final note: the way we defined a fuzzy exists fails to take into account any consideration of the way we sample the space $X$ or more precisely $X^\N$. 
Fortunately this is easy to incorporate in the Markov kernel framework and in fact makes things simpler. 
Choose a probability distribution  $\P:  \star \to X^\N$, i.e. a sampling strategy. Then the composition
\begin{equation} 
	\star \xrightarrow{\P} X^\N \xrightarrow{P_\N} \B^\N \xrightarrow{\OR_\N} \B
\end{equation}
is a probability distribution on $\B$,  which gives 
a single number $p_\P$ for the probability of $\true$
\begin{align}
	p_\P &= \int_{X^\N} (\OR_{\N*}P_\N)(\true|(x_1, x_2, \ldots)\P(d(x_1, x_2, \ldots)) \\
	         &= \sum_{(a_1, a_2\ldots) , \logicor_i a_i = \true } \int_{X^\N} P_\N((a_1, a_2, \ldots)| (x_1, x_2, \ldots)\P(d(x_1,x_2, \ldots)
\end{align}     
This number has a different interpretation than existence: it is the expected confidence to find a tuple $(x_1, x_2, \ldots)$ 
such that the predicate $P$ is true for at least one $x_i$ \emph{given our sampling strategy} $\P$.  
It does gives a lower bound
\begin{equation}
	p_\P \le \esssup_\P  (\OR_{\N*}P_\N)(\true|(x_1,  \ldots)) \le \sup (\OR_{\N*}P_\N)(\true|(x_1, \ldots) = \pfuzzy_{\exists x:P(x)}
\end{equation}
and may well be more useful than $\pfuzzy_{\exists x:P(x)}$ for many purposes. 

 \section{Conclusion}
 We have described fuzzy logic as a functorial``shadow'' of ordinary boolean logic. All the subtlety comes from the applying boolean logic not to to boolean predicates but to 
boolean valued Markov kernels (Markov predicates) that express the confidence in the truth of a predicate.  However this means that logic formulas become sensitive to the joint confidence in the predicates. We have described how this works out in the simple cases of a binary logic relation and arbitrary finite relations and simple infinte ones like the existential quantifier.  

\section{acknowledgements}
The author is grateful for useful discussions with prof. Uzay Kaymak, and dr. Hans Onvlee.


\begin{thebibliography}{LSBDW}
\bibitem[ATV]{AlsinaTrillasValverde1983}
  \bibauthor{Alsina, Claudi and Trillas, Enric and Valverde, Lloren{\c{c}}},
  \bibtitle{On some logical connectives for fuzzy sets theory},
  \bibpublication{Journal of Mathematical Analysis and Applications},
  \bibvolume{93 (1)},
  \bibpages{15--26},
  \bibpublisher{Academic Press}
  \bibyear{1983},
  \bibavailableat{https://core.ac.uk/download/pdf/82410529.pdf}

\bibitem[DP1]{Dubois-Prade1989}
  \bibauthor{Dubois, Didier and Prade, Henri},
  \bibtitle{Fuzzy sets, probability and measurements},
  \bibpublication{European Journal of Operational Research}
  \bibvolume{40}
  \bibpages{135 --154} 
  \bibpublisher{North Holland}
  \bibyear{1989}
  \bibavailableat{https://www.sciencedirect.com/science/article/pii/0377221789903263}

\bibitem[DP2]{Dubois-Prade1993}
  \bibauthor{Dubois, Didier and Prade, Henri},
  \bibtitle{Fuzzy sets and probability: misunderstandings, bridges and gaps},
  \bibpublication{Second IEEE International Conference on Fuzzy Systems},
  \bibpages{1059--1068},
  \bibpublisher{IEEE}
  \bibyear{1993},
  \bibavailableat{https://ieeexplore.ieee.org/stamp/stamp.jsp?tp=\&arnumber=327367}
 

\bibitem[Do]{Doob1984}
  \bibauthor{Doob, Joseph L},
  \bibtitle{Classical potential theory and its probabilistic counterpart},
  \bibpublication{Grundlehren der Mathematischen Wissenschaften},
  \bibvolume{262},
  \bibpublisher{Springer}
  \bibyear{1984}
  
 \bibitem[Ga]{Gaines1978}
  \bibauthor{Gaines, Brian R},
  \bibtitle{Fuzzy and probability uncertainty logics},
  \bibpublication{Information and Control},
  \bibvolume{38 (2)},
  \bibpages{154--169},
  \bibpublisher{Elsevier}
  \bibyear{1978}

  
 \bibitem[Gi]{Giry1982}
   \bibauthor{Giry, Mich\`elle},
   \bibtitle{A categorical approach to proability theory},
   \bibpublication{Categorical aspects of topology and analysis},
   \bibpages{68--85},
   \bibpublisher{Springer},
   \bibyear{1982}  
   \bibavailableat{https://link.springer.com/content/pdf/10.1007/BFb0092872.pdf}
   


\bibitem[La]{Lawvere1962}
  \bibauthor{Lawvere, F.W.}
  \bibtitle{The category of probabilistic mappings}
  \bibpublication{Seminar handout with notes by Gian Carlo Rota.}
  \bibyear{1962}
  \bibavailableat{https://ncatlab.org/nlab/files/lawvereprobability1962.pdf}


\bibitem[LK]{LiuKerre1998}
  \bibauthor{Liu, Yaxin and Kerre, Etienne E},
  \bibtitle{An overview of fuzzy quantifiers.(I). Interpretations},
  \bibpublication{Fuzzy Sets and Systems},
  \bibvolume{95 (1)},
  \bibpages{1--21},
  \bibpublisher{Elsevier}
  \bibyear{1998},


\bibitem[Lo]{Loginov1966}
  \bibtitle{Probability treatment of Zadeh membership functions and their use in pattern recognition},
  \bibauthor{Loginov, Vasiliy I},
  \bibpublication{Engineering Cybernetics},
  \bibvolume{2},
  \bibpages{68},
  \bibpublisher{SCRIPTA TECHNICA-JOHN WILEY \& SONS 605 THIRD AVE, NEW YORK, NY 10158}
  \bibyear{1966}

\bibitem[LSBDW]{lavioletteSeamanBarettDouglasWoodall1995}
  \bibauthor{Laviolette, Michael and Seaman, John W. and Barrett, J Douglas. and Woodall, William H.},
  \bibtitle{A probabilistic and statistical view of fuzzy methods},
  \bibpublication{Technometrics},
  \bibvolume{37 (3)},
  \bibpages{249--261},
  \bibpublisher{Taylor \& Francis}
  \bibyear{1995},
  \bibavailableat{https://www.tandfonline.com/doi/abs/10.1080/00401706.1995.10484327}



\bibitem[McL]{MacLane2013}
  \bibauthor{Mac Lane, Saunders},
  \bibtitle{Categories for the working mathematician},
  \bibpublication{Graduate Texts in Mathematics}
  \bibvolume{5},
  \bibpublisher={Springer Science \& Business Media}
  \bibyear{2013}


\bibitem[Pa]{Panangaden1999}
  \bibauthor{Panangaden, Prakash},
  \bibtitle{The category of Markov kernels},
  \bibpublication{Electronic Notes in Theoretical Computer Science},
  \bibvolume{22},
  \bibpages{171--187},
  \bibpublisher{Elsevier}
  \bibyear{1999},
  \bibavailableat{https://www.sciencedirect.com/science/article/pii/S1571066105806024, \hfil
  https://web.archive.org/web/20190419000326id\_/https://core.ac.uk/download/pdf/82421775.pdf}


\bibitem[Ru]{Rudin1970}
 \bibauthor{Rudin, Walter},
 \bibtitle{Real and Complex Analysis P. 2},
 \bibpublisher{McGraw-Hill}
 \bibyear{1970},




\bibitem[Vo]{Voevodsky2008}
\bibauthor{Voevodsky, Vladimir},
\bibtitle{Notes on categorical probability},
\bibyear{2008},

 
 \bibitem[Za]{Zadeh1995}
  \bibauthor{Zadeh, Lotfi A},
  \bibtitle{Discussion: Probability theory and fuzzy logic are complementary rather than competitive},
  \bibpublication{Technometrics},
  \bibvolume{37 (3)},
  \bibpages{271--276},
  \bibpublisher{Taylor \& Francis}
  \bibyear{1995}
 
 \end{thebibliography}
 \end{document}